\documentclass[
preprint,
amsmath,amssymb,
aps,
pre
]{revtex4-1}

\usepackage{amstext}
\usepackage{amsmath}
\usepackage{amssymb}
\usepackage{graphicx}
\usepackage{color} 
\usepackage{dcolumn}
\usepackage{bm}

\def\Vec#1{\mbox{\boldmath $#1$}}

\begin{document}


\title{Wavelet-based regularization of the Galerkin truncated 
three-dimensional incompressible Euler flows 
}

\author{Marie Farge}
\affiliation{%
CNRS--INSMI, LMD--IPSL, Ecole Normale Sup\'erieure--PSL,
~24 rue Lhomond,~75231~Paris~Cedex~05,~France
}%

\author{Naoya Okamoto}
\affiliation{
Center for Computational Science, Nagoya University, Nagoya, 464-8603, Japan
}
\author{Kai Schneider}
\affiliation{
I2M-CNRS, Centre de Math\'ematiques et d'Informatique,
Aix-Marseille Universit\'e, 39 rue F. Joliot-Curie, 13453 Marseille Cedex 13, France
}%
\author{Katsunori Yoshimatsu}
\affiliation{
Institute of Materials and Systems for Sustainability, Nagoya University, Nagoya, 464-8603, Japan
}

\date{\today}

\begin{abstract}
We present numerical simulations of the three-dimensional Galerkin truncated incompressible Euler equations that we integrate in time
while regularizing the solution by applying a wavelet-based denoising. For this, at each time step, the vorticity filed is decomposed
into wavelet coefficients, that are split into strong and weak coefficients, before reconstructing them in physical space to obtain the corresponding coherent and incoherent vorticities.
Both components  are  multiscale and orthogonal to each other. Then, by using the Biot--Savart kernel, one obtains the coherent and incoherent velocities. Advancing the coherent flow in time, while filtering out the noise-like incoherent flow, 
 models turbulent dissipation and corresponds to an adaptive regularization.
In order to track the flow evolution in both space and scale, a safety zone 
is added in wavelet coefficient space to 
the coherent wavelet coefficients. It is shown that the coherent flow indeed exhibits  
an intermittent nonlinear dynamics and a $k^{-5/3}$ energy spectrum, where $k$ is the wavenumber, characteristic of { three-dimensional homogeneous isotropic turbulence}.
Finally, we compare the dynamical and  statistical properties of Euler flows subjected to four kinds of regularizations: 
dissipative (Navier--Stokes), hyperdissipative (iterated Laplacian), dispersive (Euler--Voigt)  and   wavelet-based regularizations.

\begin{description}
\item[PACS numbers]
47.27.E-,
47.27.Gs,
47.27.er
\end{description}
\end{abstract}

\pacs{Valid PACS appear here}
\maketitle

\section{Introduction}
A major challenge in computational fluid dynamics is the numerical simulation of  high Reynolds number turbulence and in particular the numerical solution of the three-dimensional (3D) incompressible Euler equations.
The nonlinearity of Euler equations excites smaller and smaller scales and the same holds for Navier--Stokes equations in the inviscid limit, which corresponds to very strong turbulence when the Reynolds number tends to infinity.
%
%
%
%
Since numerical schemes are limited to a finite number of modes, or grid points, the numerical integration of Euler equations requires to apply some kind of regularization to obtain a physically relevant solution for a given  resolution.
Ideally such techniques should preserve the flow's nonlinear dynamics and the solution's properties.
For example, vortex methods introduce a cut-off in the Biot--Savart kernel. 
In the context of finite volume/difference methods, typically upwind techniques are used which introduce numerical diffusion and also numerical dispersion.
Spectral methods have the advantage to avoid numerical diffusion and dispersion, and furthermore they do preserve the conservation properties of the governing equations.
Truncated Fourier Galerkin approximations used to solve Euler equations conserve kinetic energy and it was shown that the solutions thus obtained tend in the limit of long time to energy equipartition between all Fourier modes, which corresponds to an isotropic energy spectrum with a $k^2$ behavior in three dimensions, where $k$ is the wavenumber \cite{Lee52}.  
For transient time numerical simulations of the  3D Euler equations integrated with a truncated Fourier Galerkin method exhibit  a $k^{-5/3}$  scaling, while at later time a  $k^2$ spectrum builds up which corresponds to the predicted energy equipartition \cite{CBD05}.
The statistics of the velocity field behave as a Gaussian white noise which satisfies the incompressibility constraint.
To obtain a physically relevant solution, typically hyperdissipative (also known as hyperviscous) regularizations are applied, which correspond to a Laplace operator which is iterated a certain number of times, as introduced in \cite{Lady62,Lion69} and applied in, e.g., \cite{W84, FS89}. 
Compared to viscous dissipation, which corresponds to the Laplace operator, much wider inertial ranges can thus be obtained for a given numerical resolution and therefore are frequently used to simulate geophysical and astrophysical flows.
Viscous and hyperviscous regularizations give rise to bottlenecks in the compensated energy spectrum, $k^{5/3} E(k)$, which become more pronounced as the order of the hyperdissipation,  (corresponding to the number of iterations of the Laplace operator) is increased \cite{LCP05}. 
Hyperviscous regularizations for 3D homogeneous and isotropic turbulent flows have been studied in \cite{SMC12}, and detailed analyses of bottleneck effects have  been published  in \cite{FKP08}. 
An inviscid regularization called Euler--Voigt model has been introduced  by Oskolkov~\cite{Osko82,CLT06}.
{ 
This regularization is of dispersive nature, which means that Fourier modes of different wavelength are no more propagated with the same group and phase velocity. 
Dispersion thus affects the phase of the Fourier modes, while diffusion modifies their amplitude.
The Euler--Voigt model can be obtained in the context of Navier-Stokes alpha models, i.e., a Helmholtz filter is applied to the momentum equation and the resulting equation is known as a simplified Bardina turbulence model. Setting the viscosity equal to zero yields the Euler--Voigt equations which formally correspond to adding the term $\alpha^2 \partial_t \Delta u$ to the momentum equation, where $\alpha > 0$ is a length scale that represents the width
of the spatial filter, see, e.g., the discussion in \cite{LaTi10}.
}

Wavelet techniques for simulating turbulent flows have been introduced in \cite{FSK99, SF00, FS01}. 
For reviews we refer to Farge \cite{Fa92}, Schneider \& Vasilyev \cite{SV10} and Farge \& Schneider \cite{FaSc15} .
Wavelet-based regularization of the one-dimensional Burgers equation and  two-dimensional incompressible Euler equations using Fourier Galerkin schemes has been presented in \cite{NFS09,PNFS13}. 
There it was shown that removing  noise in the truncated Fourier Galerkin simulations of the inviscid equations does yield results similar to the viscous equations. 
Applying coherent vorticity extraction, introduced in  \cite{FS01,FSK99}, to high Reynolds number 3D turbulence shows that the incoherent velocity field exhibits indeed an energy spectrum with a $k^2$ slope \cite{FPS01, OYSFK07}. 
This wavelet-based extraction method presents the advantage over the linear Fourier \cite{FSPWR03} and the nonlinear Fourier \cite{YOFS10} filtering method for extracting coherent structures out of turbulent flows. 
These previous studies motivate the present work.

{ The aim of this study is the application of wavelet-based regularization to the truncated Fourier Galerkin approximation of Euler equations to examine if wavelet-based denoising would yield the resulting flows which have similar properties as Navier--Stokes flows in the fully-developed turbulent regime. 
The idea is to remove the  noise corresponding to the $k^2$ spectrum and to check  if this is equivalent to modeling turbulent dissipation as already suggested in \cite{FPS01, SFPR05}. 
We also compare the results obtained using wavelet-based denoising with several other kinds of regularization of Euler equations, including hyperdissipative regularization by iterated Laplacian, and dispersive regularization based on the Euler--Voigt model \cite{Osko82,CLT06}.} 

The outline of the paper is the following. In Section II, we describe the governing equations, the different regularization methods of the Euler equations used here, and the numerical schemes to implement them.
In Section III, we present the results of the numerical experiments we have performed and analyze them using several statistical diagnostics and visualizations.
In Section IV, we draw some conclusions and propose perspectives for future work.

\section{Euler equations and regularization methods}
First, we describe the Euler equations and the numerical methods  used to solve them.
Then, we introduce a wavelet-based regularization of the Euler equations, and present two more classical methods, one dissipative and one dispersive, in order to compare the regularized Euler solutions thus obtained.

\subsection{Euler equations and numerical method}
We consider a  velocity field $\bm u ({\bm x},t)$
obeying the  3D incompressible Euler equations, 
\begin{eqnarray}
\partial_t {\bm u} + ({\bm u} \cdot \nabla) {\bm u} + \nabla p &=& {\bm 0}, \label{eq:mom} \\ \quad
\nabla \cdot {\bm u} &=& 0,  \quad  \label{eq:divnul}
\end{eqnarray} 
for ${\bm x}=(x^1,x^2,x^3) $ in a periodic box $\Omega = [0, 2\pi]^3$, where unit density is assumed.
The pressure is denoted by  $p(\bm{x},t)$,  $\partial_t \equiv \partial / \partial t$  and $\nabla \equiv (\partial/\partial {x^1}, /\partial{x^2}, \partial/\partial{x^3})$.
We omit the arguments $\Vec{x}$ and $t$, unless otherwise stated.

A truncated Fourier Galerkin approximation of the Euler equations   (\ref{eq:mom}) and (\ref{eq:divnul}) is obtained by developing the velocity field and the pressure into truncated Fourier series, e.g., ${\bm u}({\bm x}, t) = \sum_{\bm k}  {\widehat{\bm u}}({\bm k},t) e^{i {\bm k} \cdot {\bm x}}$, and requiring that the weighted residual vanishes with respect to test functions, which are identical to the trial functions $e^{i {\bm k} \cdot {\bm x}}$. 
Here ${\bm k}=(k^1,k^2,k^3)$ is the wave vector and $i = \sqrt{-1}$.
The incompressibility constraint is taken into account by eliminating pressure, which yields the Euler equations in Fourier space; $\partial_t {\widehat u}^\ell({\bm k})=-P_{\ell m} {\widehat N}^m({\bm k})$, where $P_{\ell m}=\delta_{\ell m}-k^\ell k^m/k^2$ and ${\bm N}=({\bm u} \cdot \nabla){\bm u}$.
Without loss of generality, we set the mean velocity $\langle \Vec{u} \rangle=\Vec{0}$, where $\langle \cdot \rangle$ denotes spatial average over the periodic box.
Then  Eqs.  (\ref{eq:mom}) and (\ref{eq:divnul}) are discretized with $N=2^{3J}=512^3 \quad (J=9)$ grid points.
The nonlinear term is evaluated with a pseudo-spectral technique, i.e., in physical space, and the aliasing errors are removed by means of the phase shift method.
Only modes with wavenumbers satisfying $k<k_{\rm max}=2^{1/2}N^{1/3}/3$
are retained.
For time integration we employ an explicit Runge--Kutta scheme of fourth order.
The dealiased pseudo-spectral discretization is equivalent to the Galerkin approximation, which by construction  does conserve kinetic energy, i.e., $d E/dt = 0$, where $E =  \int_{\Omega} |{\bm u}|^2 d{\bm x}/2$.

\subsection{Regularization methods}

\subsection*{Wavelet-based regularization}
After a brief description of the orthogonal wavelet decomposition (i) and the  nonlinear wavelet filtering (ii), we describe the procedure of   wavelet-based regularization (iii).
  The choice of the threshold used in (iii) is described (iv).
  The wavelet-based denoising regularization depends on the solution projected onto an orthogonal wavelet basis and is therefore adaptive. Since some wavelet coefficients are discarded, it has a dissipative effect. In order to obtain statistically stationary states, a solenoidal forcing term $\Vec{f}$ is imposed.

(i) {\it \underline {Orthogonal wavelet decomposition}}\\
The 3D orthogonal wavelet transform unfolds a $2\pi$-periodic vector field ${\bm v}({\bm x}, t)$ 
at a given instant $t$ into scale, positions and seven directions ($\mu = 1, ..., 7$) using a 3D mother wavelet $\psi_\mu({\bm x})$, which is based on a tensor product construction.
The wavelet $\psi$ is well-localized in space $\Vec x$, oscillating, and smooth.
The mother wavelet generates a family of wavelets $\psi_{\mu,\lambda} (\Vec x)$ by dilation and translation, which yields an orthogonal basis of $L^2({\mathbb{R}}^3)$, { and also of $L^2({\mathbb{T}}^3)$ with $\mathbb{T} = 2 \pi \mathbb{R} / {\mathbb{Z}}$ being the torus}
through the application of a periodization technique \cite{Mallat98}.
The spatial average of $\psi_{\mu,} (\Vec x)$, denoted by $\langle \psi_{\mu,\lambda} \rangle$, vanishes for each index.
The multi-index $\lambda= (j,i_1, i_2, i_3)$ denotes  the scale $2^{-j}$ and  position $2\pi \times 2^{-j} {\bm i} =2\pi \times 2^{-j}(i_1, i_2, i_3)$ of the wavelets for each direction.

A vector field $\bm{v}(\bm x)=(v^1,v^2,v^3)$ sampled on $N=2^{3J}$ equidistant grid points, having  zero mean value, can be decomposed into an orthogonal wavelet series: 
\begin{equation}
{\bm v}  ({\bm  x}) = 
\sum_{j=0}^{J-1} {\bm v}_j({\bm  x}), 
\label{OWS}
\end{equation}
where ${\bm v}_j$ is the contribution of ${\bm v}$ at scale $2^{-j}$ defined by 
\begin{equation}
{\bm v}_j({\bm  x}) = \sum_{\mu=1}^7 \sum_{i_1,i_2,i_3 =0}^{2^j -1} {\widetilde {\bm v}}_{\mu, \lambda}  \psi_{\mu, \lambda} (\Vec x).
\label{OWS_2}
\end{equation}
Due to orthogonality of the wavelets, the coefficients are given by $ {\Vec {\widetilde v}}_{\mu,\lambda}  = \left< {\Vec v}, \psi_{\mu,\lambda} \right>$, where $\left<\cdot,\cdot\right>$ denotes the $L^2$-inner product defined by $\left< f, g \right> = \int_{\Omega} f(\Vec x) \, g(\Vec x) d\Vec x$. 
At scale $2^{-j}$ we have $N_j=7 \times 2^{3j}$ wavelet coefficients for each component of $\Vec{v}$.
Thus, in total we have  $N$ coefficients for each component of the vector field corresponding to $N-1$ 
wavelet coefficients and the vanishing mean value. 
These coefficients are efficiently computed from the $N$ grid point values for each component of $\Vec v$ using the fast wavelet transform, which has linear computational complexity.
In the present work, the compactly supported Coiflet wavelets with filter width 12 are used.
For more details on wavelets, we refer the reader to text books, e.g., Mallat \cite{Mallat98}.

(ii) {\it \underline { Wavelet-based denoising}}\\
Thresholding the wavelet coefficients ${\widetilde {\Vec{v}}}_{\mu,\lambda}$ at a given time instant, we can define the coherent subset of the  wavelet coefficients ${\widetilde {\Vec{v}}}_{\mu,\lambda}^{\rm c}$ by
\begin{equation}
{\widetilde {\Vec{v}}}_{\mu,\lambda}^{\rm c}
= \left\{
\begin{array}{lll}
{\widetilde {\Vec{v}}}_{\mu,\lambda} & {\mathrm{ for }} & |{\widetilde {\Vec{v}}}_{\mu,\lambda}|> {T} , \\ 
0 & {\mathrm{ for }} & |{\widetilde {\Vec{v}}}_{\mu,\lambda}| \leq {T},
\end{array}
\right. 
\label{thresholding_fct}
\end{equation}
where $T$ is a given threshold value. The choice of the threshold value is discussed in (iv).  
The coherent field $\Vec{v}_c$ is then reconstructed by  inverse wavelet transform.
The remaining incoherent field $\Vec{v}_i$ is given as $\Vec{v}_i=\Vec{v}-\Vec{v}_c$.

(iii) {\it \underline {Wavelet-based regularization of Euler equations}}\\
The numerical simulation of the Euler equations with wavelet-based regularization is also called Coherent Vorticity Simulation (CVS).
The procedure of CVS, starting from the Fourier coefficients of the velocity field $\Vec{{\hat u}} (\Vec{k},t)$ at $t=t_n$, is as follows.\\
(a) {\it  Time integration in spectral space:} 
The velocity $\Vec{{\hat u}} (\Vec{k},t)$ is advanced in time in $\Vec{k}$--space up to $t=t_{n+1}$ using the fourth-order Runge--Kutta method.\\
(b) {\it  Reconstruction of vorticity in physical space:} \\
The vorticity field $\Vec{\omega}=\nabla \times \Vec{u}$ at $t= t_{n+1}$ is reconstructed by applying 
the inverse Fourier transform to $\Vec{{\widehat \omega}}=i \Vec{k} \times \Vec{{\hat u}}$.\\
(c) {\it  Extraction of coherent vorticity and addition of safety zone in wavelet space:} \\
The set of wavelet coefficients of vorticity is obtained by applying the fast wavelet transform to $\Vec{\omega}$.
In order to track the evolution of coherent vorticity in space, scale and direction we have to keep, not only the coherent wavelet coefficients ${\widetilde {\Vec{\omega}}}_c$, but also the neighboring wavelet coefficients in space, scale and directions.
For this we first define the index set $\Lambda$ that is the union of all $(\mu, \lambda)$ corresponding to the coherent wavelet coefficients kept in Eq. (\ref{thresholding_fct}). 
We then define an expanded index set $\Lambda_*$ which adds to $\Lambda$ the indices $(\mu, \lambda)$ of the neighboring coefficients in position, scale and direction. 
For  details on the definition of the safety zone, we refer to \cite{OYSFK11}. 
Finally, all the coefficients, which do not belong to $\Lambda_*$ are set to zero.
 The expanded wavelet coefficients indexed by $\Lambda_*$ correspond to the coherent ones plus those of the safety zone, and are denoted by ${\widetilde {\Vec{\omega}}}_{c*}$.\\
(d) {\it  Reconstruction in physical space of the expanded coherent vorticity:} \\
Applying the inverse wavelet transform to ${\widetilde {\Vec{\omega}}}_{c*}$ yields the coherent vorticity including the safety zone $\Vec{\omega}_{c*}$.\\
(e) {\it  Calculation of the expanded coherent velocity:} \\
The induced velocity $\Vec{u}_{c*}$, which is divergence free, is computed using the Biot--Savart relation 
$\Vec{u}_{c*}=-\Delta^{-1} (\nabla \times \Vec{\omega}_{c*})$ in wavenumber space.
The steps (a)--(e) are  applied in each time step.

(iv) {\it \underline {Choice of the threshold}}\\
The choice of the threshold is motivated by the fact that  CVS of Euler equations does not work for threshold values $T=0$ or $T= \infty$. This is because  the former corresponds to the simulation for Euler equations   (all wavelet coefficients are kept) and thus the $k^2$ range grows with time, while   no  coefficients are retained in the latter CVS. 
It is anticipated that there are appropriate values such that CVS can simulate flows without a $k^2$ range in the energy spectrum.
After some trial and error to avoid the appearance of the $k^2$ range, we selected the value $T=2T_0$, where $T_0=\{(4/3)Z\ln N \}^{1/2}$ and the enstrophy $Z=\langle |\bm {\omega}_{c*}|^2\rangle/2$. 
 The value of $T_0$ is based on the Donoho threshold without iteration used in \cite{OYSFK11} to perform CVS of Navier--Stokes equations. \\
 
\subsection*{Other regularizations}
For the sake of comparison, we will consider other kinds of regularization of the Euler equations. 
We will thus add a term having a dissipative effect on the solution, and another one having a dispersive effect, to obtain the regularized Euler equations 
\begin{equation}
\partial_t \bm{u} + (\bm{u}\cdot \nabla) \bm{u}
+\nabla p=
  \nu_h (-1)^{h+1} \nabla ^{2h} \bm{u} 
  +\alpha^2 \nabla^2 \partial_t \bm{u} + \bm{f}, \quad \quad
\nabla \cdot \bm{u} =0,
\label{eq:EV}
\end{equation}
where $\nu_h (-1)^{h+1} \nabla ^{2h} \bm{u}$ is a dissipative term,
$\alpha^2 \nabla^2 \partial_t \bm{u}$ is a dispersive term,
and $\bm{f}$ is a solenoidal forcing term. 

{\underline {\it Dissipative regularization}}\\
We study two kinds of dissipative regularizations.
The dissipative term in Eq. (\ref{eq:EV}) has a non-zero positive coefficient, i.e., $\nu_h > 0$, while
the dispersive term  vanishes, i.e., $\alpha=0$.
Since energy is then dissipated we need to add a forcing term $\bm{ f}$ in Eq. (\ref{eq:EV}) to keep the flow statistically steady.\\
 {(a) \it Viscous regularization (Navier--Stokes equations)}\\
The choice $h=1$  results in the regular Newtonian viscosity term and Eq. (\ref{eq:EV}) then corresponds to the Navier--Stokes equations.\\
 {(b) \it Hyperviscous regularization}\\
Higher integer values of $h$ 
correspond to different kinds of hyperdissipation, for which the energy dissipation becomes more and more localized in a narrower and narrower range of high wavenumbers in Fourier space, see, e.g., \cite{SMC12}.
{This implies a longer inertial range at the expense of the dissipation range which is thus reduced.} 

{\underline{\it Dispersive regularization}}\\
We also perform a simulation where we apply the Euler--Voigt regularization (EV)   to the Euler equations \cite{CLT06,Osko82}. 
In this case, $\nu_h$ and $\Vec{f}$ are set to zero in Eq. (\ref{eq:EV}).
Since it is an inviscid regularization, whose effect is dispersive rather than dissipative, the modified energy, defined as $E_m = E + \alpha^2 Z$, is conserved in time.
The Euler--Voigt regularization with parameter $\alpha = 0$ corresponds to the Euler equations and, since we solve them using a Fourier Galerkin scheme, energy cascades and piles up at the cutoff wavenumber during the flow evolution. 
To avoid such a pile-up, the value of $\alpha$ has to be sufficiently large and here we  choose $\alpha$ = 2/5.

\subsection{Numerical methods  used for the regularizations}
In total we have performed five flow simulations:
\begin{itemize}
\item[i)] Euler equations with wavelet--based denoising (CVS),
\item[ii)] Navier--Stokes equations (NS),
\item[iii)] Euler equations with hyperviscous regularization (HV),
\item[iv)]  Euler equations with dispersive regularization (EV),
\item[v)] Euler equations without any regularization (Euler).
\end{itemize} 

We apply the same Fourier Galerkin method to discretize in space all those governing equations. For the time integration we use a fourth order Runge--Kutta method, with the time increment chosen as $1.0 \times 10^{-3}$ to insure that the Courant--Friedrichs--Lewy number remains below $0.4$ for all the computations.
For the Euler equations, with and without Euler--Voigt regularization, we set $\nu_h=0$. Therefore we do not need to add any forcing (${\bm f}={\bm 0}$), since either kinetic energy $E$ or modified energy $E + \alpha^2 Z$ is conserved.
For the remaining  computations, we add a solenoidal random forcing $\Vec{f}$, whose time correlation is $1.0 \times 10^{-3}$ and  with magnitude $1.0 \times 10^{-3}$, to compensate the dissipated energy and obtain  a statistically stationary state.
The forcing ${\bm f}$ is applied only in the low wavenumber range $1 \le k<2.5$.
Readers interested in details on generating  such a random force are referred to \cite{OYSFK11}. 
Note that we use the same realization of the random forcing in all simulations.
For the coefficients of the dissipative regularizations, we set $\nu_1=4.0\times 10^{-4}$ 
for the Navier--Stokes equations, and $\nu_4=1.5\times 10^{-14}$ for the hyperviscous regularization with $h=4$.
The choice of these coefficients is determined in such a way that the enstrophy, obtained when the flow evolution has become quasi-steady, has about the same value for the different simulations made with either dissipative regularizations or CVS regularization. 
For the Euler--Voigt regularization, we set $\nu_h=0$ and $\alpha=2/5$. 

For all simulations, except the one with hyperviscous regularization (HV), the number of  grid points $N$  is $512^3$.
For HV we use $N_{\rm HV}=256^3$ and only those Fourier modes with wavenumbers  smaller than the cutoff wavenumber $k_c$ are retained.
This choice is motivated by the fact that the number of retained modes matches the number of retained wavelet coefficients in CVS.
 Thus $k_{ c}$ is set to $105$ using $4\pi k_c^3/3\sim N_{\rm CVS} $, where $N_{\rm CVS}=0.036\times 512^3$, as we shall see later.

\begin{table}[tb]
  \begin{tabular}{l|cccc}
    Type   & $\nu_h$   & $\alpha$                & wavelet   &  \quad forcing term      \\
	\hline
    CVS   & 0          & 0                  & Yes  & \quad Yes        \\
    NS   & $\nu_1=4.0\times 10^{-4} $ & 0                  & --   & \quad Yes        \\
    HV   & $\nu_4=1.5\times 10^{-14}$ & 0                  & --   & \quad Yes        \\
    EV    & 0          & $\alpha=2/5$         & --   & \quad --      \\
	Euler    & 0          & 0                  & --   & \quad --        
  \end{tabular}
  \caption{
  Summary of the different regularization methods presented  in this article, mentioning the dissipative parameter $\nu_h$ and the dispersive parameter $\alpha$ which have been used.  }
  \label{reg}
\end{table}
 A summary of the different regularizations compared here, indicating the parameters used for each simulation, is given in Table \ref{reg}.
The initial condition of all simulations, except HV, corresponds to a fully developed turbulent flow at 
 Taylor microscale Reynolds number $R_\lambda=257$, which was obtained by a direct numerical simulation forced with a negative viscosity as explained in \cite{IKYIU07}.
For the simulation with hyperviscous regularization (HV), we use an initial velocity field which retains only the modes whose wavenumber is below $k_c$.

{Concerning the CPU time, CVS is about 60 \% more expensive than NS, EV, HV and EE, which are about the same. In the current implementation the wavelet transform is not optimized and its parallelization is based on a transposition technique, which requires global data communication and hence slows down the computation. For CVS the ultimate goal is to perform Euler simulations directly in an adaptive wavelet basis, thus reducing memory and CPU time requirements. Viscous dissipation is then absent and dissipation is only due to filtering out the incoherent part. In Roussel and Schneider \cite{RoSc10} computations of a slightly compressible turbulent mixing layer showed a speed-up of the computation and memory reduction for CVS of about a factor $3$ in comparison to DNS of Navier--Stokes. The fully adaptive version advances in time only the coherent flow (represented by few wavelet coefficients), and adds a safety zone at each time step to account for translation of vortices and the generation of finer scales. 
We anticipate that similar performance will be obtained in a fully adaptive version of the CVS Euler code.
}

\section{Numerical Results}
\label{sec:result}
In the following we discuss the results obtained for the five 
flow simulations.

\subsection{Time evolution of  statistics}

\begin{figure}[tb]
\begin{center}
\includegraphics[width=0.5\linewidth,clip]{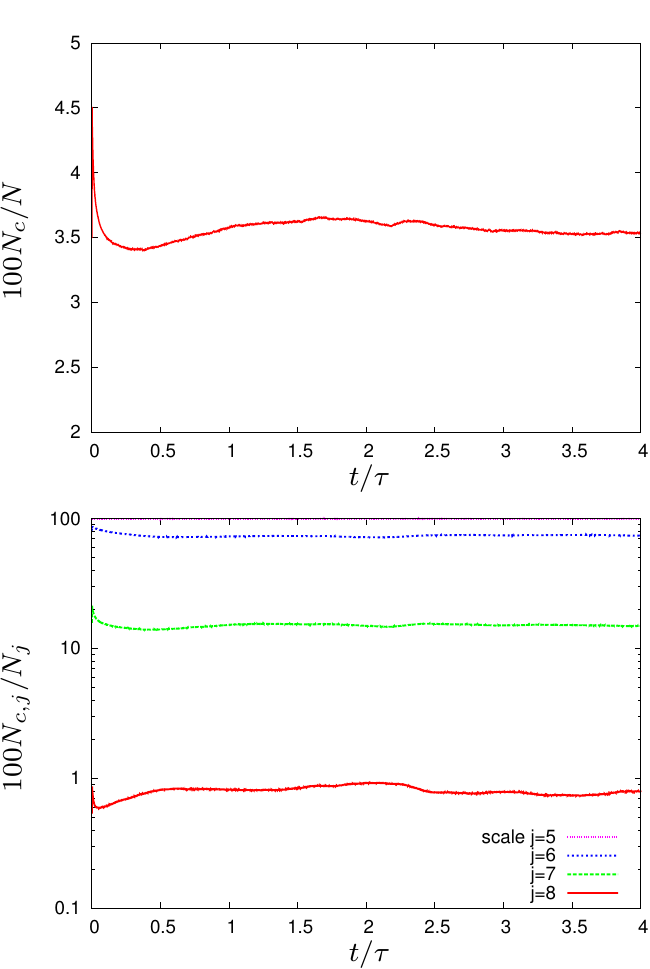}
\end{center}
\vspace{0.5cm}
\caption{
Time evolution of the percentage of
the retained
wavelet coefficients, $100N_c/N$ (top),
and  the percentage of
 the retained 
wavelet coefficients
at each scale $j$, $100N_{c,j}/N_j$ (bottom).
\label{DoF}}
\end{figure}

All   computations are integrated in time for about four initial eddy turnover times $\tau=L/u_{0}$, where $u_0=\sqrt{ 2E/3}$ and $L$ is the integral length scale defined by $L=\pi/(2u_0^2)\int_0^{k_{\rm max}} e(k)/k dk$, with $e(k)$ being the isotropic energy spectrum, defined as $e(k) =\frac{1}{2} \sum_{k-1/2 \leq |\bm{p}| < k+1/2} |\bm{{\widehat u}}(\bm{p})|^2$.

Figure \ref{DoF} (top) shows the time evolution of the percentage of the wavelet coefficients retained by CVS, namely $100 N_{ c}/N $, where $N_{ c}$ is the number of the wavelet coefficients which correspond to the coherent flow, including the safety zone.
We see that the percentage of retained coefficients does not vary much, around $3.5\%$, after a transient decay for $t \lesssim 0.2\tau$. 
The computations presented here do not benefit from this compression in terms of  computational cost, since the flow field is reconstructed in Fourier space or in physical space on the full grid  $N=512^3$ at each time step and a spectral method is used for space discretization. 
Nevertheless, the percentage of retained wavelet coefficients remains a good indicator of the potential gain which  can be achieved by adaptive wavelet simulations \cite{FS01, SV10}.
 In Fig. \ref{DoF} (bottom), we plot the number of wavelet coefficients  that  CVS retains at each scale, namely $100 N_{c,j}/N_j $, where $N_{c,j}$ is the number of  wavelet coefficients which correspond to the coherent flow including the safety zone, at a given scale indexed by $j$.
Note that $\sum_{j=0}^{J-1} N_{c,j}=N_c$.

\begin{figure}[tb]
\begin{center}
\includegraphics[width=0.5\linewidth,clip]{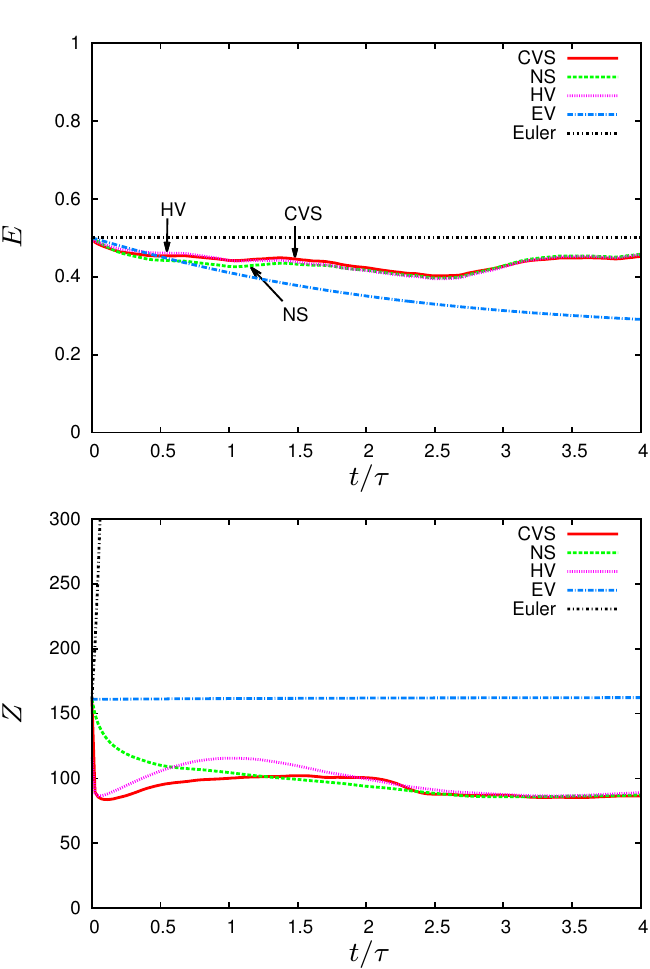}
\end{center}
\vspace{0.5cm}
\caption{Time evolution of  energy $E$ (top), and
 enstrophy $Z$ (bottom).
 \label{ene_ens}
}
\end{figure}

\begin{figure}[tb]
\begin{center}
\includegraphics[width=0.5\linewidth,clip]{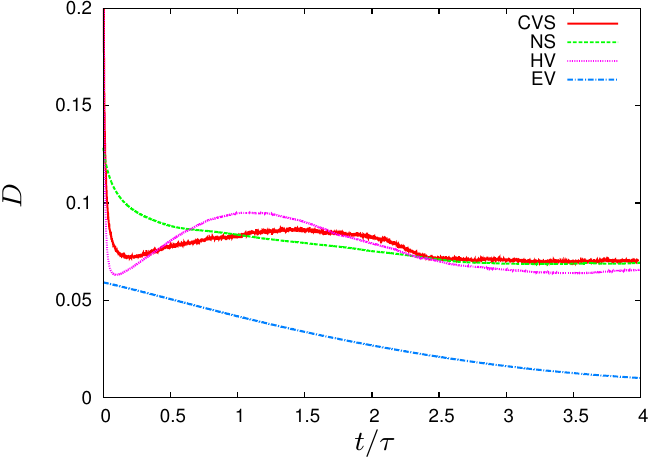}
\end{center}
\vspace{0.5cm}
\caption{Time evolution of 
the  dissipation rate $D$.
 \label{dissp}}
\end{figure}

From the largest scale $j=0$ to scale $j=5$, we observe that 100\% of the wavelet coefficients are retained as coherent flow, while at scale $j=6$ the percentage drops to $80\%$, then to $15\%$ at scale $j=7$, and finally to less than $1\%$ at the smallest scale $j=8$. 
Therefore it is the compression obtained at the smallest scales that dominates, since the number of coefficients, $N_j$, drastically increases with the scale index $j$, as $N_j = 7 \times 2^{3j}$.

Figure \ref{ene_ens}  plots the time evolution of turbulent kinetic energy, 
$
E= \langle |\bm {u}|^2 \rangle/2
$, 
and of enstrophy,
$
Z=\langle |\Vec{\omega}|^2 \rangle/2$,
 for the Euler equations and for the different regularized Euler equations, namely CVS, NS, HV, and EV.
Figure \ref{ene_ens} (top) shows 
 that both CVS 
 and HV present the same time evolution of the energy as NS.
It also confirms that the numerical scheme used to solve the Euler equations is sufficiently conservative, since only 0.024 \% of the initial energy is lost after four eddy turnover times $\tau$ which is due to the time discretization. 
 In contrast, for EV energy decreases significantly in time, because only the modified energy, $E+\alpha^2Z$, is conserved (this within $4.3 \times 10^{-6} \%$ of its initial value after $t=4 \tau$).
 Figure \ref{ene_ens} (bottom) 
 shows how, after a transient period up to $t=2.5 \tau$, enstrophy reaches almost the same value for CVS, HV and NS, as expected (since we have adjusted the parameter $\nu_h$ in HV and NS to match the level of enstrophy of CVS for the steady state).
For the Euler case, enstrophy grows rapidly in time due to energy piling up at high wavenumbers in absence of regularization. 
For the dispersive regularization EV, Fig. 2 (bottom) shows that the enstrophy is  almost conserved, which suggests that the nonlinear transfer of energy towards smaller scales is inhibited.
The values of energy $E$ and enstrophy $Z$  at $t=3.4\tau$ are summarized in Table  \ref{tab_1} for all computations.

Let us recall that  the nonlinear wavelet filtering of CVS regularization removes the noise-like incoherent part from the flow at each time step, and thus CVS is dissipative.
In order to estimate the energy dissipation rate $D$ in this case, we use 
 \begin{eqnarray}
 D= \langle \bm{u}\cdot \bm{f}\rangle - \frac{ dE}{dt},
 \label{DD}
 \end{eqnarray}
where $d E/dt$ is estimated by the first order  forward finite difference in time. 
Since NS and HV have  dissipative terms, energy dissipation can be directly estimated by
$\langle \epsilon_h \rangle=\nu_h (-1)^h \langle \Vec{u}\cdot \nabla^{2h} \Vec{u} \rangle$.
 We verified that  the difference between the values  estimated by the two methods, i.e., either $D$ or $\langle \epsilon_h \rangle$, is negligibly small for NS and HV  (the differences are less than $0.17\%$ for $t>0.1\tau$).
It can be noted that Eq. (\ref{DD}) is similar to what was  used to estimate numerical viscosity in \cite{DXS03}.
 
 Figure \ref{dissp}  plots the energy dissipation rate $D$ for CVS, NS, HV, and  EV, but not for the Euler equations which conserve energy.
 We observe that the energy dissipation of CVS is close to those for NS and HV  for $t > 2.5 \tau$. This suggests that the mean energy dissipation rate $D$  is insensitive to the detailed structure of the vorticity field. 
 This shows that using CVS removes the incoherent noise-like contribution to the flow which corresponds to energy dissipation. Indeed, the incoherent enstrophy is a measure of turbulent dissipation. 
The insensitivity of the energy dissipation rate is consistent with the observation in \cite{YAK15} that the scrambling the high wavenumber contribution of the flow field does not modify the mean energy dissipation rate.
The value of $D$ for EV is determined by $D=-dE/dt$, because $\bm{f}=\bm{0}$.

\begin{table}[tb]
  \begin{tabular}{l|cccccc}
    Run   & $E$   & \quad $Z$                & $ \quad A$   & $ \quad L $ & \quad $\lambda$ & \quad $R_\lambda^S$   \\
	\hline
    CVS   & 0.448 & \quad 85.53             & \quad  0.46  & \quad　1.07  & \quad　0.162     & \quad　217      \\
    NS   & 0.453 & \quad 86.22              & \quad  0.46  & \quad　1.11  & \quad　0.162     & \quad　226      \\
    HV   & 0.450 & \quad 86.57              & \quad  0.43  & \quad　1.11  & \quad　0.161     & \quad　228      \\
    EV   & 0.303 & \quad 162                & \quad  0.10  & \quad　0.657 & \quad　0.097     & \quad　--       \\
	Euler   & 0.500 & \quad 1.41$\times$$10^4$ &  \quad  0   & \quad　0.260 & \quad　0.013      & \quad　--       
  \end{tabular}
  \caption{
  Energy $E$, enstrophy $Z$, normalized energy dissipation rate $A=DL/u_0^3$, integral scale $L$, Taylor microscale $\lambda$ and the Taylor microscale Reynolds number $R_\lambda^S$
 at $t=3.4\tau$.  
  The statistics for CVS, NS and HV are statistically stationary, while the non-conservative statistics of EV and Euler are time-dependent.}
  \label{tab_1}
\end{table}

The normalized mean energy dissipation rate $A=DL/{u_0}^3$ is a key quantity 
to study the phenomenology of turbulence.
Our results listed in Table \ref{tab_1} show that the values of $A$ are slightly smaller than 0.5 for CVS, NS, and HV.
These values  agree excellently  with  asymptotic values for isotropic turbulence at high Reynolds number ($R_\lambda \gtrsim 200$) obtained by DNS of the Navier-Stokes equations \cite{KIY03} and hyperviscous computations \cite{HB04}.
In \cite{SMC12},  the Taylor-microscale Reynolds number as a function of $L/\lambda$, $R_\lambda^S=36.4L/\lambda-23.1$, was introduced using data fitting and applied for hyperviscous computations, where $\lambda$ is the Taylor-microscale $\lambda=\sqrt{5E/Z}$.
The values of $R_\lambda^S$  for our dissipative regularizations (HV and NS) are also summarized in Table \ref{tab_1}.
The value of $R_\lambda^S$ for NS is 226 at $t=3.4\tau$, which  is close to the value of 223 estimated by the classical definition of $R_\lambda=u_0\lambda/\nu$ for NS.
The value of $R_\lambda^S$ for CVS is $R_\lambda^S=217$, which is close to the values of $R_\lambda^S$ for NS and HV.
The values of  the corresponding Taylor-microscale $\lambda$  are also listed in Table \ref{tab_1}  and we find very similar values for CVS, NS and HV.

\subsection{Energy spectra and fluxes}
\begin{figure}[tbp]
\begin{center}
\includegraphics[width=0.45\linewidth,clip]{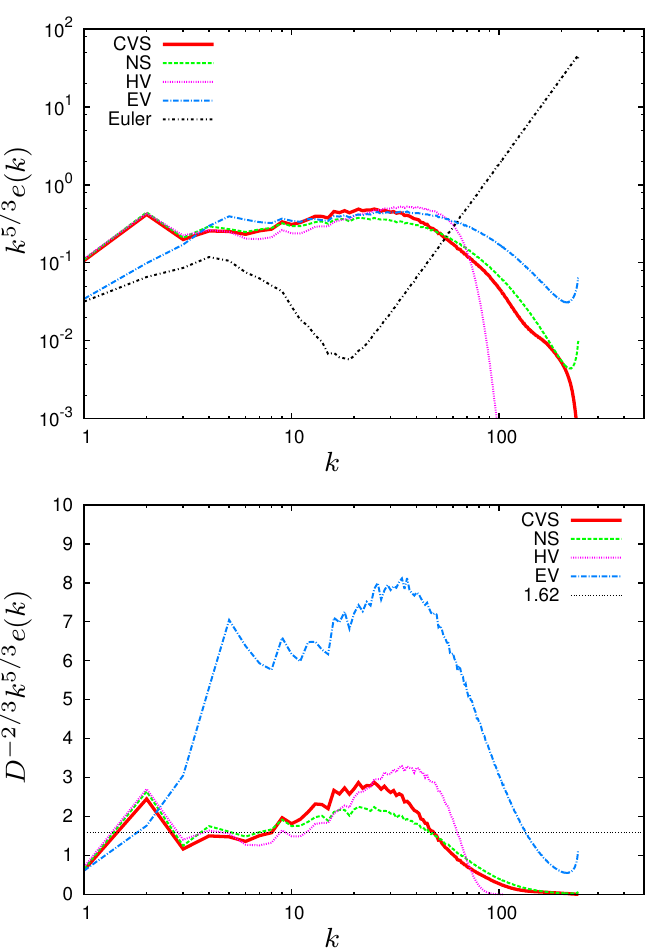}
\end{center}
\vspace{0.5cm}
\caption{ Compensated energy spectra $k^{5/3}e(k)$ in log-log plot (top), 
and $D^{-2/3}k^{5/3}e(k)$ in semi-log plot (bottom) at $t=3.4\tau$.
\label{enespe}
}
\end{figure}

\begin{figure}[tbp]
\begin{center}
\vspace{-0.1cm}
\includegraphics[width=0.5\linewidth,clip]{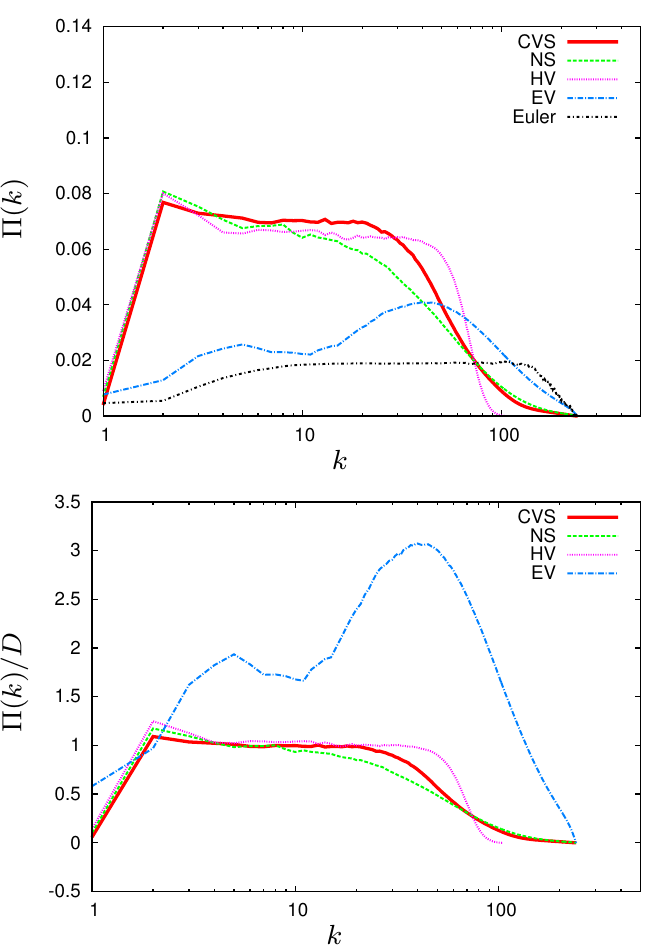}
\end{center}
\vspace{0.5cm}
\caption{ Energy flux $\Pi(k)$ (top), and 
  $\Pi(k)/D$ (bottom) at $t=3.4\tau$. \label{flaspe}}
\end{figure}

{ 
To get insight into the spectral distribution of turbulent  kinetic energy, we plot in Fig. \ref{enespe} the compensated energy spectrum $k^{5/3}e(k)$ for the five flows as a function of wavenumber $k$ at time $t= 3.4 \tau$.
In both the energy containing and the inertial range ($k \lesssim 10$) we observe that  CVS, NS and HV yield similar compensated energy spectra.
In contrast for EV it  substantially differs from the others due to the absence of large scale forcing.
At moderate wavenumbers ($10 \lesssim k \lesssim 60 $) we find for CVS and NS similar spectral behaviors.
We also notice that the compensated spectra of all regularizations, including NS, exhibit bottlenecks with different peak wavenumbers $k_p$ ($k_p=20$  for CVS and NS, $k_p=35$ for HV and EV).
For large wavenumbers ($k > 60$) the energy spectrum is significantly damped for HV compared to NS due to the hyper-dissipative term.
Moreover, CVS retains much more energy than HV and a little less than NS as the noise removed by CVS is predominant at high wavenumbers, due to its $k^2$ behavior.
Now considering the Euler case, we find that for all wavenumbers the energy spectrum differs from the four other cases and, in particular, we observe that $e(k) \propto k^2$ for $k > 20$, which corresponds to energy equipartition. 
Notice, to compute the 1D energy spectrum $e(k)$ we have integrated the 3D energy spectrum over spherical shells of radius $k$, i.e., the shell surface scales as $k^2$. Hence, energy equipartition in 3D Fourier space corresponds to $e(k) \propto k^2$.
Figure \ref{enespe} (bottom) shows the compensated energy spectrum nondimensionalized by the energy dissipation rate $D$, namely $D^{-2/3}k^{5/3}e(k)$. 
In the inertial range ($3 \lesssim k \lesssim 10$)  we observe 
that both CVS and HV keep almost constant values, similar to NS,
which are close to the value $1.62$ assumed for the Kolmogorov constant.
The value $1.62$ for the 3D energy spectrum is obtained by applying the
correction factor $55/18$ \cite{MoYa75} to the value $0.530$ of the Kolmogorov constant estimated from a large set of experimental data in \cite{Sr95} for the 1D longitudinal energy spectrum.

Figure \ref{flaspe} (top) plots the energy fluxes $\Pi(k)$ for CVS, NS, HV, EV and Euler.
Here,   $\Pi (k)$ and $T(k)$ are defined by $\Pi (k) = -\int_0^k T(x) dx$ and $T(k) = -\sum_{k-{1}/{2} \le |{\bm p}| < k + {1}/{2}} \Vec{{\hat u}}(-{{\bm p}}) \cdot \Vec{{\widehat N}}({\bm p})$, respectively, where $\Vec{N}=(\Vec{u}\cdot \nabla) \Vec{u}$.
We observe  that in the inertial range ($k \lesssim 10$) the energy flux $\Pi (k)$ of CVS is close to those of NS and HV, while for moderate wavenumbers ($10 < k \lesssim 60$) $\Pi (k)$ of CVS and HV are more pronounced than the energy flux of NS.
This suggests that CVS and HV well preserve the nonlinear dynamics of turbulence in the inertial range. 
The flux $\Pi(k)$ in EV and Euler is significantly reduced for $k\lesssim 10$ compared to the fluxes of CVS, NS and HV, because forcing is absent in EV and Euler.
Note that $\Pi(k_{\rm max})=0$ due to the solenoidal constraint of the velocity and the skew symmetry of the nonlinear term $T(k)$.
}

{The energy fluxes  normalized by the energy dissipation rate $D$ are shown in Fig. \ref{flaspe} (bottom) excluding the Euler case.
We observe that NS, CVS and HV exhibit a plateau range, where $\Pi(k)/D \sim 1$, for $k >2$.
For NS the plateau ends at $k \sim 10$, corresponding to the end of the intertial range.  
As expected, HV exhibits the longest plateau up to $k \sim 40$, extending the inertial range at the expense of the dissipative range, which is thus reduced.
The plateau of CVS ends at $k \sim 20$ and the corresponding energy flux remains in between NS and HV. 
Indeed, CVS offers a kind of interpolation between NS and HV.
In contrast, for EV no plateau is observed and maximum energy flux is found at $k=50$ being three times larger than for the other cases, showing that EV is very different from NS.}

\subsection{ Visualizations and $Q-R$ diagrams}

\begin{figure}[h!]
\begin{center}
\includegraphics[width=0.7\linewidth,clip]{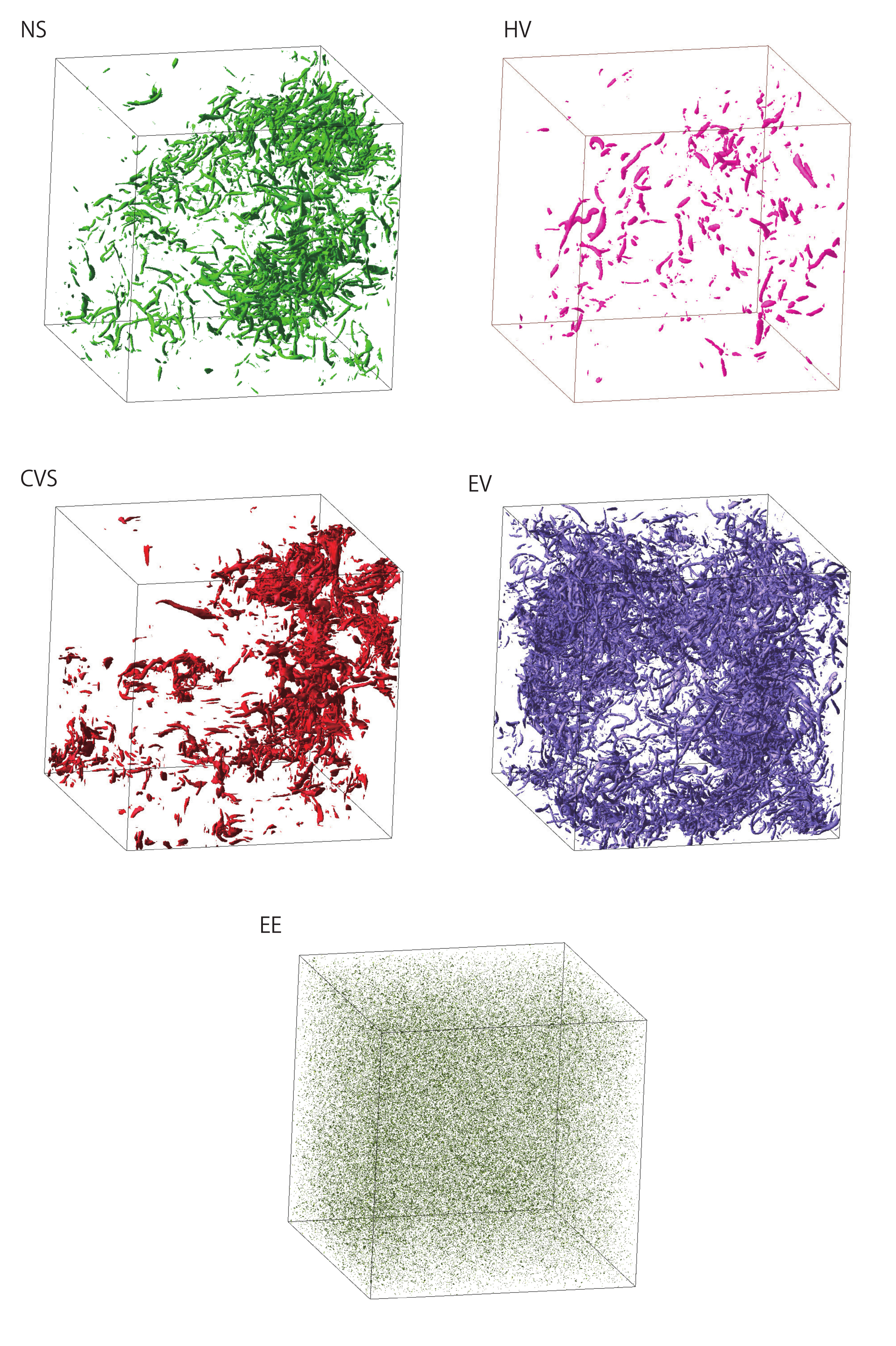}
\end{center}
\vspace{0.5cm}
\caption{ Visualization of  intense vorticity regions for 
NS (green), CVS (red),  HV (magenta), EV (purple)
and Euler (gray)
at $t=3.4\tau$.
Isosurfaces of vorticity are shown for
$|\bm{\omega}|=M+4\sigma$, where
$M$ and $\sigma$ denote respectively the mean value
and standard deviation of the modulus of the vorticity field
of NS.
The  values of $M$ and $\sigma$ 
 are
10.2 and 8.27, respectively.
Only $1/8$ subcubes  are shown to enlarge the structures.\label{visualvor}}
\end{figure}

Figure \ref{visualvor} shows the most intense structures of the vorticity field for CVS, NS, HV, EV and Euler, visualized by the  isosurface $|\bm{\omega}|= M + 4 \sigma$, $M$ being the mean value and $\sigma$ the standard deviation of the modulus of vorticity for the Navier--Stokes simulation.
The isosurface value of $|\bm{\omega}|$ is the same for all computations.
We observe that vorticity structures are tube-like for CVS, NS, HV and EV.
The structures of HV are more sparsely distributed compared to NS which is consistent with \cite{SMC12}.
{In contrast, we do not see any coherent structures in the Euler solution, which behaves as a Gaussian white noise since the $k^2$ scaling of the energy spectrum (Fig. \ref{enespe}) corresponds to decorrelation in physical space
and the PDF of the longitudianl velocity derivative is Gaussian (Fig. \ref{pdfsD}).
}

{We also analyzed the velocity gradient tensor $\partial u^i/\partial x^j$ (for a review we refer to \cite{Wall09}) of the five different flows at $t= 3.4 \tau$. We study the second and and third invariants, $Q = -\frac{1}{2} \partial u^i/\partial x^j \partial u^j/\partial x^i$ and $R = -\frac{1}{3} \partial u^i/\partial x^j \partial u^j/\partial x^\ell \partial u^\ell/\partial x^i$, respectively, as proposed in \cite{ChPC90}.
The joint PDFs of the dimensionless invariants, called $Q-R$ diagram, in Fig. ~\ref{fig_qr} present a very similar teardrop shape for CVS, HV and EV, close to the shape found for NS. In contrast, for EE we observe a symmetric joint PDF with respect to the line $R=0$, which exhibits a keyhole shape.
These observations illustrate that the small scale properties of the CVS, HV and EV flows agree well with those observed for NS, which is not the case for what we find for EE. 

\begin{figure}[htb]
\begin{center}
\includegraphics[width=0.4\linewidth,clip]{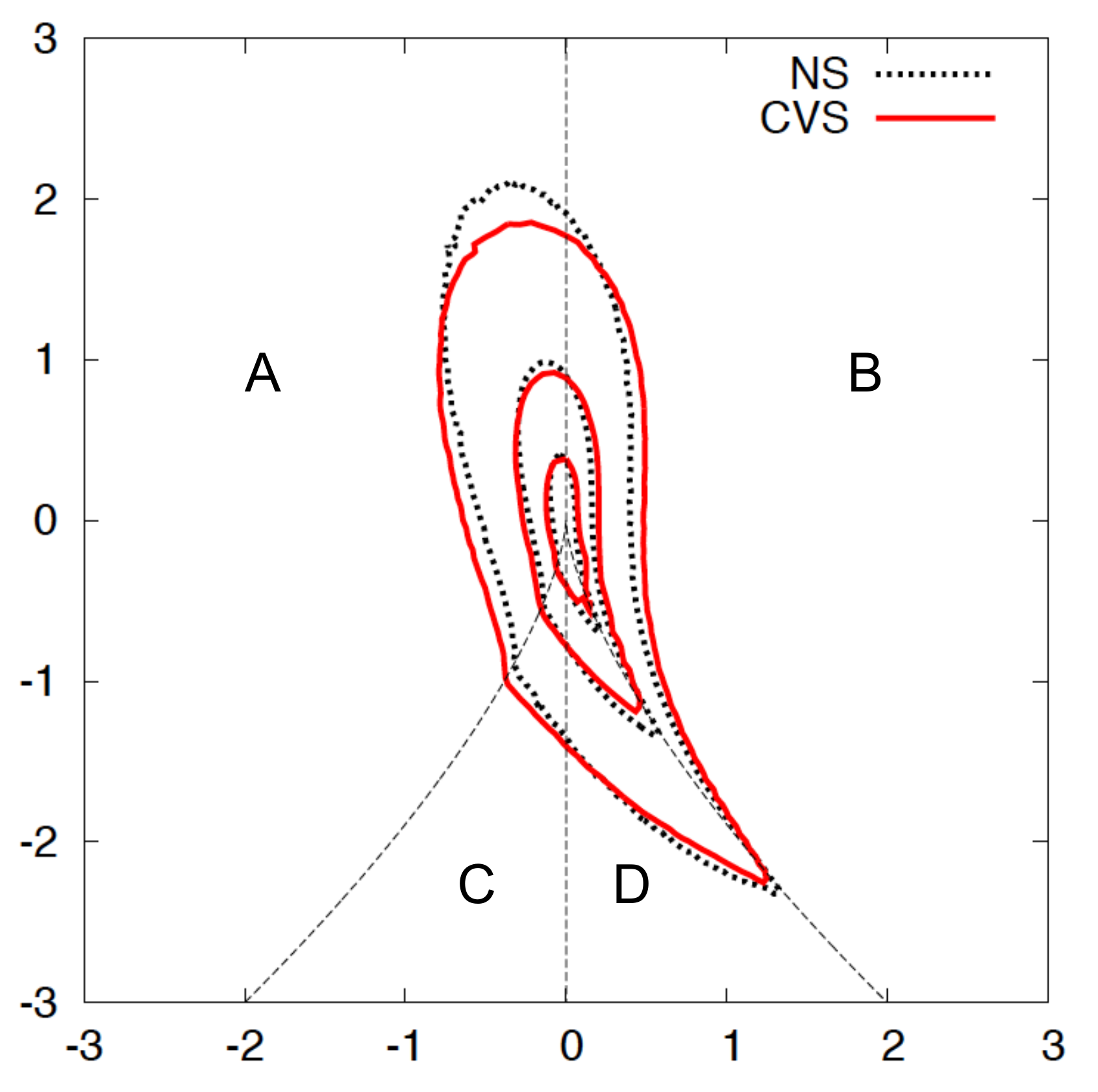}
\includegraphics[width=0.4\linewidth,clip]{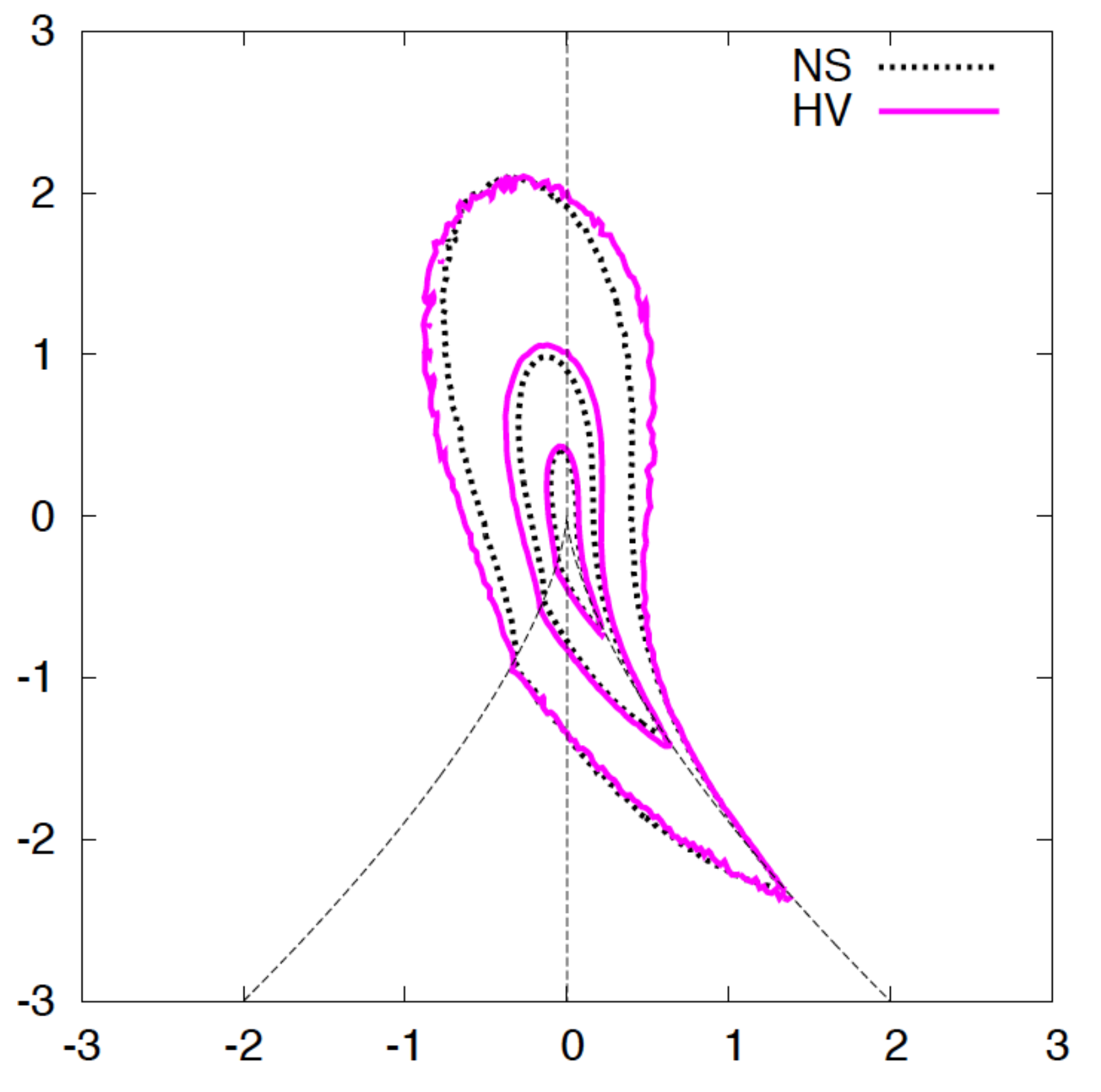}\\
\includegraphics[width=0.4\linewidth,clip]{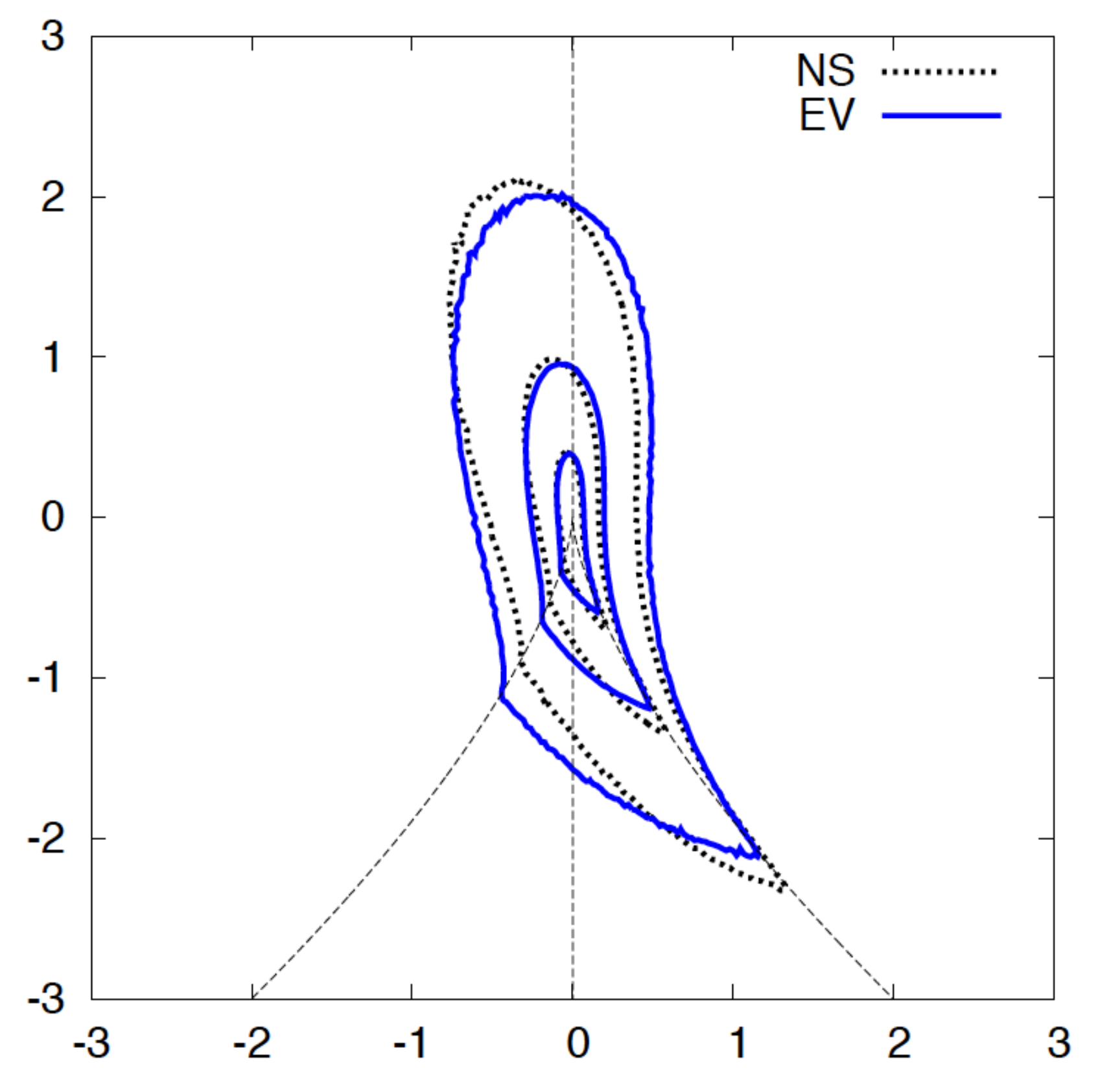}
\includegraphics[width=0.4\linewidth,clip]{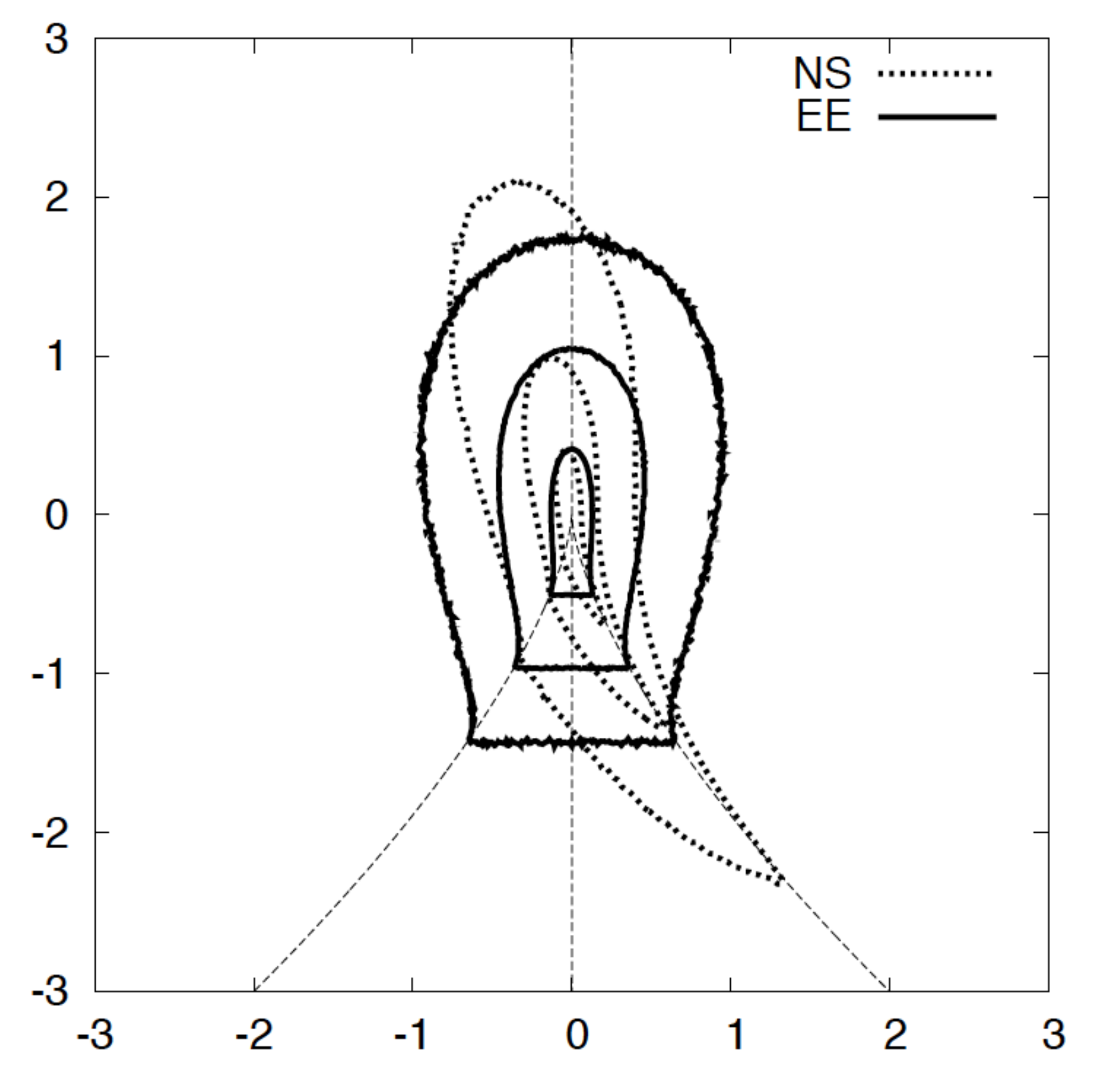}
\end{center}
\vspace{0.5cm}
\caption{{Joint PDFs of the dimensionless invariants $Q/ \langle S_{ij} S_{ij} \rangle$
and $R/ \langle S_{ij} S_{ij} \rangle^{3/2}$ where $S_{ij} = \frac{1}{2}\left (\partial u^i/\partial x^j + \partial u^j/\partial x^i \right)$ for CVS, HV, EV and EE in comparison with NS (dotted lines). Shown are the isolines $1, 10^{-1}$ and $10^{-2}$ together with the so-called Vieillefosse curve \cite{Vieil84} given by $27 R^2 / 4 + Q^3 = 0$ which distinguishes the four domains corresponding to different local flow patterns (A: stable focus-stetching, B: unstable focus-compressing, C: stable node-saddle-saddle, D: unstable node-saddle-saddle) and the vertical line $R=0$ (dashed lines).}
 \label{fig_qr} }
\end{figure}
}

\subsection{Probability density functions and scale-dependent flatness}
\begin{figure}[htb]
\begin{center}
\includegraphics[width=0.5\linewidth,clip]{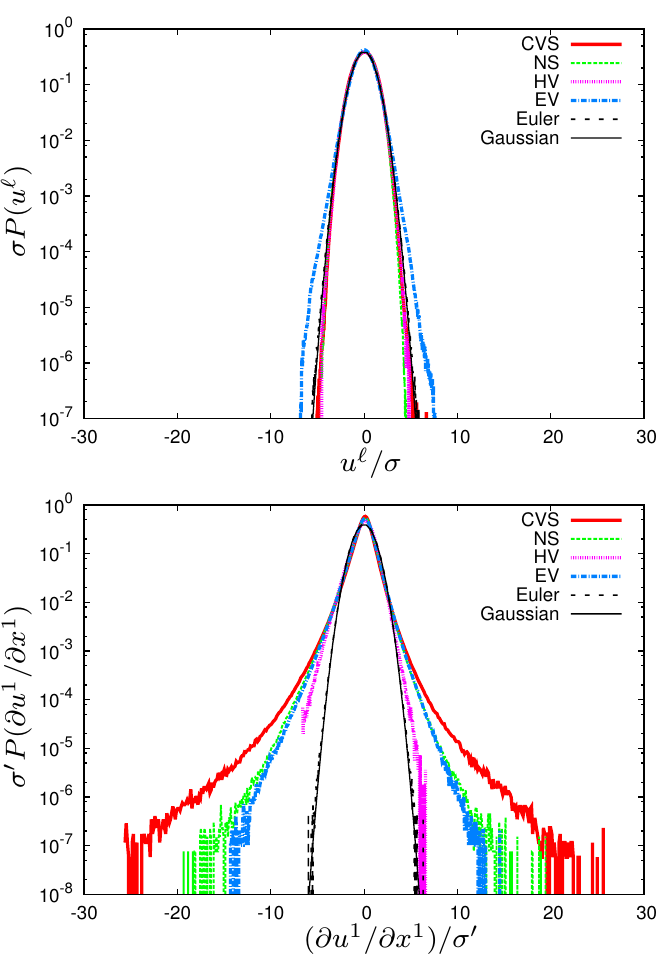}
\end{center}
\vspace{0.5cm}
\caption{PDFs of  velocity (top), and
 longitudinal velocity derivative $\partial u^1/\partial x^1$ (bottom) at $t=3.4\tau$. 
 The Gaussian distribution is
  plotted as a reference.
 \label{pdfsD} }
\end{figure}

{
Now we  show  in Fig. \ref{pdfsD} the probability density functions (PDFs) of velocity, $P[u^{\ell}]$, and of the longitudinal velocity derivative, $P[\partial u^1 / \partial x^1]$, estimated using their histograms computed with 200 bins. Each PDF is normalized by its standard deviation.
We observe that the shape of the velocity PDF for each case, except for EV, remains close to
the shape of the normal distribution.
Table \ref{tab_2} summarizes skewness and flatness factors for $u^\ell$: The skewness values $S[u^\ell]$ are small and negative (of order $10^{-2} \sim 10^{-1}$) for all computations.
The flatness values $F[u^\ell]$ for CVS, NS, HV and Euler are close to $3$, the flatness of the normal  distribution, while for EV it is $3.52$, which confirms a slight departure from Gaussianity.

The longitudinal velocity derivative $\partial u^1/\partial x^1$ is a quantity well suited to characterize small scale intermittency. 
In the following we study its PDF, $P[\partial u^1/\partial x^1]$,  for the five simulations as shown in Fig. \ref{pdfsD} (bottom).
Each PDF is again normalized by the corresponding standard deviation.
First we find that the skewness of the longitudinal velocity derivative $S[\partial u^1/\partial x^1]$ is negative for all cases, and for CVS its value is closer to the one of NS compared to the other cases, as shown in Table \ref{tab_2}.
For the Euler case this skewness almost vanishes, which confirms its Gaussian behavior.
While for the Euler case the PDF is indeed Gaussian (with flatness $F[\partial u^1/\partial x^1] = 3.02$), the other cases progressively depart from Gaussianity, reflected in heavier tails, this in the order: HV ($F[\partial u^1/\partial x^1] =4.30$), EV ($6.32$), NS ($6.90$) and CVS ($11.2$), as given in Table \ref{tab_2}.
These findings confirm that CVS is more intermittent than NS, while HV is less intermittent,
the latter being consistent with previous work \cite{SMC12}.
For the Euler case the normal distribution proves that the flow is non intermittent. 
}

\begin{table}[tb]
  \begin{tabular}{l|cccc}
    Run   & $S[u^\ell]$ & \quad $S[\partial u^1/\partial x^1]$ & \quad $F[u^\ell]$  & \quad $F[\partial u^1/\partial x^1]$  \\
	\hline
    CVS   & $-5.4\times 10^{-2}$  & -0.60  & 2.82   & 11.2  \\
    NS   & -0.12                 & -0.54  & 2.84   & 6.90  \\
    HV   & $-5.8\times 10^{-2}$  & -0.43  & 2.86   & 4.30  \\
    EV   & $-3.1\times 10^{-2}$  & -0.38  & 3.52   & 6.32  \\
	Euler   & $-1.1\times 10^{-2}$  & $-3.7\times 10^{-4}$  & 3.02     & 3.02  
  \end{tabular}
  \caption{Skewness and flatness factors 
  of velocity and longitudinal velocity derivative at $t=3.4\tau$. }
  \label{tab_2}
\end{table}

\begin{figure}[tb]
\begin{center}
\includegraphics[width=0.5\linewidth,clip]{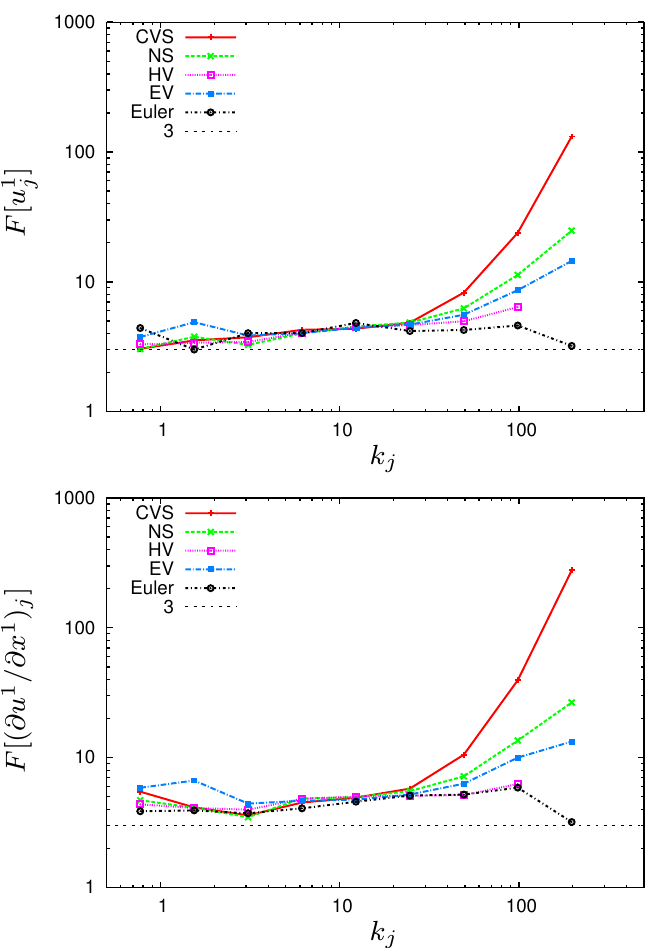}
\end{center}
\caption{Scale-dependent flatness factors for $u^1$ and
 $\partial u^1/\partial x^1$
 at time $t=3.4\tau$.
\label{sbs_flat}}
\end{figure}

Wavelet coefficients allow us to further quantify the flow intermittency \cite{BLS09,YOSKF09}, since wavelets are well-localized functions in space, which are contracted and dilated to explore a large range of scales.
The scale-dependent flatness  at scale $j$ is defined by the flatness of a wavelet-filtered quantity.
The scale index $j$ corresponds to the wavenumber $k_j=k_\psi 2^j$,  where $k_\psi$ is the centroid wavenumber of the chosen wavelet  ($k_\psi=0.77$ for the Coiflet 12 used here).
For a wavelet-filtered quantity at  $k_j$, $v^\ell_j(\Vec{x})$, given by Eq. 
(\ref{OWS_2}), we define the scale-dependent flatness of $v^\ell_j(\Vec{x})$ by 
\begin{equation}
F[v^\ell_j]=\langle ( v^\ell_j   )^4 \rangle/
\langle ( v^\ell_j   )^2 \rangle.
\label{mom}
\end{equation}
(Note that $\langle v^\ell_j \rangle=0$.) Figure \ref{sbs_flat} plots the scale-dependent flatness for the $x^1$-component of velocity, $F[u^1_j]$, and for the longitudinal velocity derivative, $F[(\partial u^1/\partial x^1)_j]$.
{
For CVS, NS and EV we observe that both $F[u^1_j]$ and $F[(\partial u^1/\partial x^1)_j]$ increase with  $k_j$,
and this is more significant for CVS after $k  \gtrsim 50$, than for NS, and less significant for EV compared to NS for the same scales.
The flatness values $F[u^1_j]$ and $F[(\partial u^1/\partial x^1)_j]$ of Euler and HV hardly depend on scale,
which shows that the two flows are not, or much less intermittent, respectively.
In contrast, the wavelet-based regularization leads to a stronger intermittency for CVS than for NS, since CVS extracts coherent structures by denoising the vorticity field at each time step.
Reversely for EV regularization intermittency of the flow is reduced with respect to NS.
The HV regularization suppresses  the flow intermittency significantly,  resulting in reduced flatness values similar to those observed for Euler. This is  also reflected in reduced tails of the PDF of the longitudinal velocity derivative, a result which
is consistent with \cite{SMC12}.}

\section{Conclusions and perspectives}

We have proposed a wavelet-based approach to adaptively regularize the solution of  three-dimensional incompressible Euler equations computed with a classical Fourier Galerkin spectral method.
We  compared the wavelet-based method with three regularizations: Navier--Stokes, hyperviscous and Euler--Voigt. 
In addition we performed computations for the Euler equations without regularization.
The main findings can be summarized as follows:
First,  wavelet-based regularization (CVS), as well as hyperviscous regularization (HV), preserve the Navier--Stokes (NS) dynamics in the inertial range 
{  selecting in both cases only a reduced set of the total number of modes used for NS.}
For the wavelet regularization the flow is more intermittent than for NS, since it extracts coherent structures 
{ by removing Gaussian decorrelated  noise at each time step. 
In contrast, the flow obtained by hyperviscous regularization is less intermittent than for NS.
CVS offers a significant compression rate reducing the number of active degrees of freedom to only about 3.5\%  for the turbulent flows studied here, i.e.,  $R_{\lambda} \sim 200$. For higher $R_{\lambda}$ flows the compression rate will even be more efficient, as shown in \cite{OYSFK07} for high resolution DNS of Navier--Stokes.}
From the time evolution of energy and enstrophy of EV, it is speculated that the large-scale flow decays in time with the small-scale flow being almost frozen.
Further studies of the flow structure  and its dynamics for different  values of $\alpha$ at higher resolution for longer time computations would lead to an improved understanding of the effect of Euler-Voigt regalization \cite{GKB15}.
In conclusion, the comparison of different regularization methods of the Euler equations shows  the potential of CVS for simulating fully developed turbulence using a reduced number of degrees of freedom, while preserving the intermittency of the flow. Perspectives for future work are  adaptive simulations of turbulent flows solving the Euler equations at a reduced computational cost and applying CVS filtering for modeling turbulent dissipation.

\begin{acknowledgments}
The computations were carried out on the FX100 system at the Information Technology Center of Nagoya University. 
This work was partially supported by JSPS KAKENHI Grant Number (S)16H06339, (A)2524701 and (C)17K05139.
MF and KS acknowledge support by the French Research Federation for Fusion Studies within the framework of the European Fusion Development Agreement (EFDA).

\end{acknowledgments}

\bibliography{apssamp}

\clearpage

	


\end{document}